# Numerical Study of Wind Pressure Loads on Low Rise Buildings under different Terrain

Saidi O. Olalere [1], and Olufemi O. Alayode [2]

[1] Obafemi Awolowo University/Department of Mechanical Engineering, Ile Ife, Nigeria; Email: olaleresaidi@gmail.com
[2] Obafemi Awolowo University/Department of Mechanical Engineering, Ile Ife, Nigeria; Email: olufemialayode@yahoo.com

*Abstract*— **This is a numerical study of wind pressure loads on low rise buildings in which three different types of roofs were analyzed which are the flat, gable and circular roof at different wind speed. The numerical analysis was performed using FLUENT package based on values of k (turbulence kinetic energy) and Ɛ (dissipation rate of turbulence) based on partial differential equation. Also, flat, and shallow escarpment terrains were considered during the simulation to determine the coefficient of pressure at different wind speed for different roof types. For the shallow escarpment terrain, a flat roof was considered at different velocities and for the flat terrain, three different types of roofs are considered which are the flat, gable and circular roof. It is observed that as the wind speed increases, the coefficient of drag decreases. It also shows the effect of vortex formed at the leeward direction of the building which implies the higher the wind speed, the larger the vortex formed and the lower the building ventilation and higher the damage on the roof of the building. Based on the analysis, it is preferable to use a circular roof based on the aerodynamic characteristics of wind around building walls and roofs.**

## I. Introduction

Wind induced dispersion of pollutants in different locations depends on turbulence characteristics and velocity profile of the wind. These will in turn depend on the roughness and general configuration of the upstream terrain. Flow over low-rise building encompasses the need to monitor both internal and external unsteady pressure, wind loads on low rise building and its load paths through both structural and non-structural components.

According to (Sifton et.al. 1998) internal pressure of design wind loads on building envelope contribute a significant portion to the total design wind load depending upon the dominant opening size and location, shape of the building, surrounding conditions and other aerodynamic factors. Design wind loads on building envelope are due to a net combination of external and internal pressure (Stathopoulous et al. 2007) in which internal and external pressure measurements are also essential for assessing infiltration or exfiltration of air, moisture movement and thermal variations through building envelope which have significant influence on both the internal environment and the energy needs of building.

Accurate assessment of internal pressures is, therefore, essential both from wind loads and energy efficiency of buildings aspects. Thus, in the presence of openings, the algebraic sum of the external and internal pressure is used to assess the design wind loads on building envelope components such as walls, roofs, roof tiles, windows, and doors. Low rise building are fully immersed within the layer of aerodynamic roughness where the turbulence intensities are high (Vickery 1986) were given as the shape of the building, the spatial variation of external pressure at the opening, the geometries of the openings, the size and location of the opening with respect to the windward as well as the background porosity, ventilation opening sizes, internal volume and compartmentalization, wind direction, upstream flow turbulence intensities and flexibility of the building envelope. The fulfillment of certain conditions of opening porosity and internal volume become a reason for the formation of turbulence energy at the opening that causes the internal pressure to exceed the external pressure fluctuation (Kopp et al 2008).

A study by (Kopp et al 2008) examined the effects of opening location and sizes, background leakage, compartmentalization, roof, and vents in which the experiment shows that the external roof pressure are highly correlated with respect to time with the internal pressure with decrease in the ratio of the internal volume to the opening area leads to increase in the internal pressures for wind directions normal to the opening.
With the invention of aerodynamic in which inflow of wind through the building envelope leads to over pressurization of the internal dwelling unless there is an equivalent opening in the leeward side to relieve the pressure. So, the aerodynamic factors that govern the magnitude and direction of internal pressure in a building are fluctuation of external pressure at the openings, the upstream wind direction, size and position of opening, internal volume and compartmentalization, natural ventilation opening and leakages due to crack and outlet ducts (Holmes 2009).

The purpose of the research is to get the numerical characteristics of wind pressure loads on low rise building under different terrain since the effect of wind pressure load on structures should be emphasize because of its negative effect on the economy. Reardon and Holmes (1981) at James Cook University gave a description on research on low-rise structures in which it was concluded that:(i) Flows perpendicular to a wall, a more turbulent environment resulted in closer reattachment, more free streamline curvature and lower pressure, and (ii) For quartering flows the action of the vortices was enhanced by roof overhangs. Reardon (1997) fatigue failure on industrial and residential low-rise buildings resulted in research in metal cladding fastener failure with repeated, cyclic, gust loading. "The worst mean roof suctions, independent of direction, occur along the edges near the windward corner, but not at the corner itself".

However, most low-rise buildings are in amongst their peers, not isolated out in a field. The impact of a field of similar structures surrounding a subject structure was the topic of extensive studies in the 1980s for the fledgling solar power industry (Derickson, 1992). Low rise buildings are routinely adversely impacted by the speed-up effect caused by terrain, as noted by Davenport (1993).

The difficulties to assess wind-induced loads for low-rise buildings arise are because, "They are usually immersed within the layer of aerodynamic roughness on the earth's surface, where the turbulence intensities are high, and interference and shelter effects are important, but difficult to quantify. Roof loadings, with all the variations due to changes in geometry, are of critical importance for low-rise buildings. The highest wind loading on the surface of a low-rise structure are generally the suctions on the roof, and many structural failures are initiated there" according to Holmes (2001).

Wind pressure load has been simulated using different software. Numerical models are based on evaluation of the spatial and time dependent properties of wind-induced pressure. The time dependent loads on buildings can be determined by Large Eddy Simulation (LES) or by Direct Numerical Simulation (DNS). The calculation of the structural response to fluctuating loading is possible with models like finite element modeling. The commercial software FLUENT 6.2 was utilized for this simulation and the governing equation employed were the Reynolds Average Navier Strokes (RANS) equations, together with the k-turbulence model. The inlet, top, outlet, and two sides of the computational domain were set at different values.

## II. Literature Review

To observe abnormalities or irregularities in the behaviors of flow around a low-rise building is relevant. These irregularities have been described by different observers, researchers, scientists, engineers which depend on how well the experiment is controlled. The experiments are carried out in a standard laboratory environment under controlled and adverse operating conditions, and the results are compared, analyzed, and interpreted. In describing wind load in building, different conceptual models have been developed. So, flow over obstacles has been extensively investigated both experimentally and numerically (Cook, 1997).

According to (Kopp et.al. 2007) internal pressure can contribute a significant portion to the total design wind load on which the intensity and distribution depends on the severity of the aerodynamic factors involved and the internal pressure account for more than 50% of the wind load. Wind induced internal pressure on low rise buildings with openings such as windows and doors can form a higher proportion of the total design wind load (Holmes 2001).

The internal pressure is affected by the complex dynamics of wind and building interaction to properly design building envelopes and components from the perspective of wind resistance, water intrusion and energy performance. So, internal pressure is affected in a complex manner by opening size and location, compartmentalization, background leakage, flexibility of envelope, internal volume, and external pressure distribution at the opening wind direction (Oh et al 2008). The interaction of wind and building causes the variation of pressure more than the resistance capacity of the building envelope that could lead to failure of the building components.

Holmes (1979) conducted a study on the internal pressure fluctuation of a single opening building model using boundary layer wind tunnel to investigate the relationship between internal pressure and Helmholtz resonance. This study revealed that the internal pressure in buildings with opening responds quickly to external pressure fluctuation like a Helmholtz resonance. Then it shows that air moves in and out of the building in response to external pressure and the internal pressure fluctuation due to the compressibility effects of the air.

A boundary layer study of the behaviors of transient wind-induced internal pressure to compare the phenomenon of overshooting to peak values of steady state internal pressure fluctuations (Lucluan,1989) in which the observation shows that the steady-state peak fluctuation is higher than the transient response overshooting. In this study, the doors and windows located on the windward side cause an increase in the density of the air inside and inflation of the building as wind rushes in which result in the buildup of positive internal pressure. Therefore, the location of opening at specific part of the envelope lead to the development of significant internal pressure variation due to the interaction of wind and building which creates a region of separation and reattachment of flows depending on the size of the building and angle of attack.

A study was also conducted to investigate the transient behavior of the internal pressure due to sudden breach of opening under smooth and turbulent flow (Vickery 1994) and a study of sustained dynamic action of turbulent wind over an opening vital of imposing damage to the building (Mehta 1992). This experiment shows that the internal pressure doesn't decay with time in the case of turbulent flow in which fluctuation of the internal pressure was equivalent to that of the external pressures. It was observed that the correlated internal pressure fluctuation with that of external pressure provides a higher peak load. The effect of openings and porosity in internal pressure was examined to evaluate its influence on the internal pressure (Woods,1995) and a numerical study performed on the viability of the synchrony of formation of sudden overshoot characteristics between wind tunnel and full-scale studies (Sharma,2010). The results of the experiment show steady state theory agrees with experimental measurement of internal pressure for the case of a single opening.

An investigation of the influence of Helmholtz resonance on internal pressure in a low rise building under oblique wind flow in which the result shows that the effect of resonance at oblique flow being significant causes large fluctuation in internal pressure (Richard,2003).

Kopp et al (2008) performed an internal volume sealed wind tunnel experimental study to examine the effects of ten different opening configurations on the internal pressure of low-rise building. The results of the experiment show that the peak internal pressure strongly correlates in time with the external pressure. The internal pressure coefficient was large when there was an opening in the windward side of the building. Wall leakages acts to ease the internal pressure fluctuation, and this could basically be due to the leakage of air through the leeward and side walls that contribute to deflating the building interior (Sharma 2007).

1.0    Low Rise Building

Low rise buildings, which implies roofed low-rise structures are between 4.0m to 4.5m in height and are frequently equate with being low-cost structures. Low rise building depends on composite action and load sharing behavior within and between wall, roof and floor system for stiffness stability and strength (Foliente 1998) with low aspect ratio (ratios of their overall height to their plan dimension), shallow foundation, flexible horizontal diaphragms and are frequently constructed with several different materials of dissimilar stiffness, strength, and mass properties.

2.0    Wind Flow Topography

Wind speed effect constitute abrupt changes in the general topography located in any exposure which increased considerably by natural and man-made topography.

2.0.1    Shallow escarpment

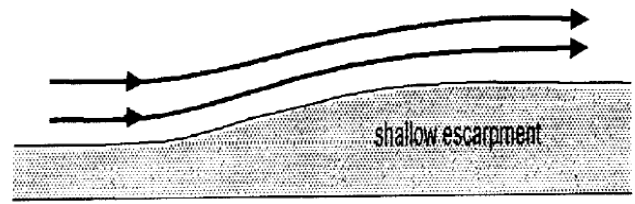

Figure 1. Shallow Escarpment (Holmes 2001)

As the wind approaches a shallow feature, its speed first reduces slightly as its encounter the start of the slope upwards as shown in Fig 1. It gradually increase in speed as it flows up the slope towards the crest. The maximum speed-up occurs at the crest or slightly upwind of it.

2.0.2    Shallow hill

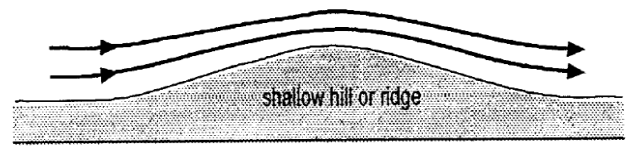

Figure 2. Shallow hill or ridge (Holmes 2001)

Beyond the crest, the flow speed gradually reduces to a value close to that well upwind of the topographic feature which is a feature with a downwind slope as shown in Fig 2.

2.0.3    Steep Escarpment

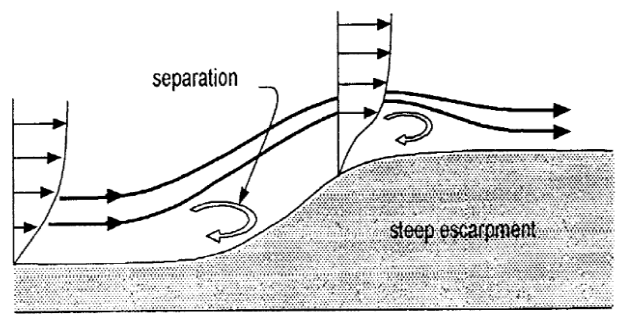

Figure 3. Steep Escarpment (Holmes 2001)

In Fig 3, the seperation occur at the start of the upwind slope, immediately downwind of the crest.

### 2.0.4 Steep Hill or Ridge

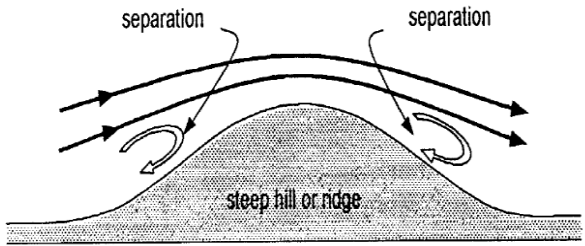

Figure 4. Steep hill or ridge (Holmes 2001)

The seperation may occur at the start of the upwind slope and on the downwind slope for a ridge as seen in Fig 4.

### 3.0 Building Roof Profile

### 3.0.1 Flat Roof

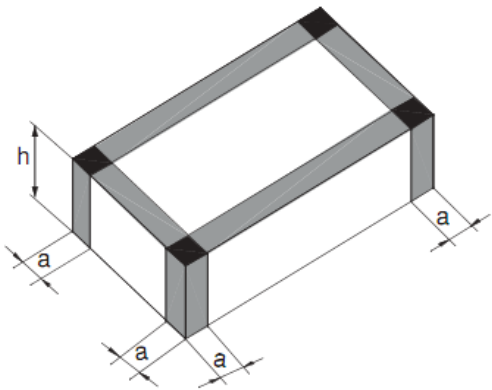

Figure 5. Flat roof (ASCE/SEI 7-05)

Fig 5 shows a sloped form of roof which is horizontal. This can be made from metal like lead (welded or folded seamed), tin (folded, soldered or folded seamed) or copper. The notation a is the horizontal dimension and h is the mean roof height.

### 3.0.2 Hip Roof

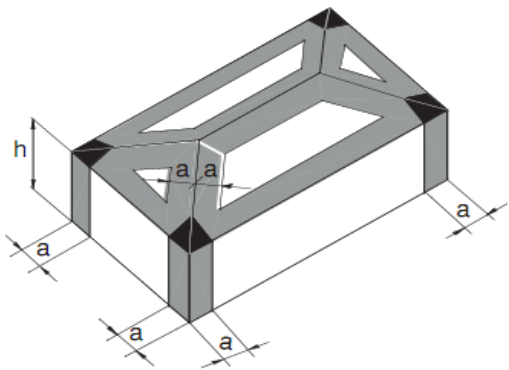

Figure 6. Hip roof (ASCE/SEI 7-05)

Fig 6 is a type of roof where all sides slope downward to the walls, usually with a fairly gentle slope.

### 3.0.3 Gable Roof

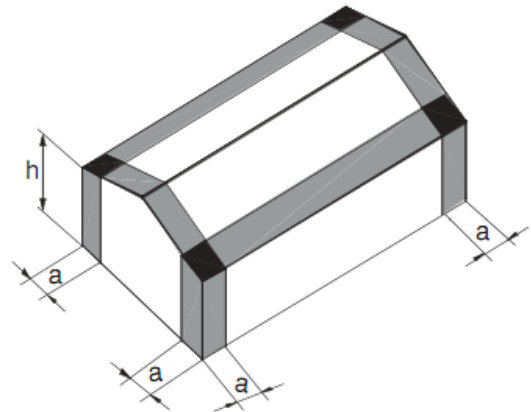

Figure 7. Gable roof (ASCE/SEI 7-05)

Fig 7 is a triangular portion of a wall between the edge of a sloping roof. The shape of the gable depends on the structural system being used.

### 4.0 Wind Flow Terrain

Selection of terrain categories are due to the effect of obstructions which constitute the ground surface roughness. The terrain category depends on the direction of wind under consideration. So, if wind in a fully developed boundary layer encounter a change of surface roughness, the adjustment starts at ground level and gradually moves upward as shown in Fig 8. The result is the development of an internal boundary layer over the new terrain (Deaves,1981). For flow from smooth terrain (roughness length $Z_1$) to rougher terrain $Z_2$ with $Z_2 > Z_1$

$$X_I(Z) = Z_2 \left(\frac{Z}{0.36 Z_2}\right)^{4/3} \quad (1)$$

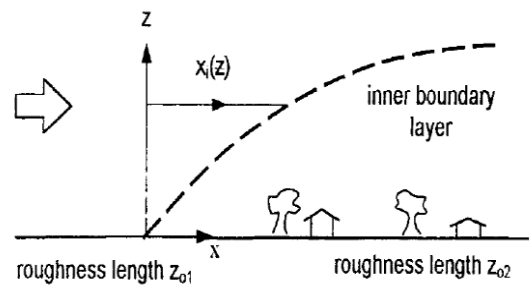

Figure 8. Internal boundary layer development at a change of terrain roughness (Holmes 2001)

Terrain in which a specific building stands can be assessed as being one of the following terrain categories;
Category 1: Exposed open terrain with a few or no obstructions and in which the average height of any object surrounding the structure is less than 1.0m.
Category 2: Open terrain with well-scattered obstructions having height generally between 3.0m and 3.5m.
Category 3: Terrain with numerous closely spaced obstructions having the size of building structure up to 4.0m in height with or without a few isolated tall structures.

Category 4: Terrain with numerous large high closely spaced obstructions.

5.0  Computational Fluid Dynamics

The rapid developments in both computer hardware and software have created a possible environment for the practical applications of CFD to simulate flows within and around buildings and other structures. Gomes (2005) presented the comparison of experiment and numerical results of wind pressure on some irregular-plan shape. Due to the complex flow field in the vicinity of a building, past investigations of CFD applications are mainly focused on rather simple geometry or low-rise buildings. The FLUENT fluid simulation software is used to simulate the wind pressure load on low rise building. The selected software is selected based on its allowance for users to create their own physical models from the user interface to develop relationships and the boundary conditions from the simulation.

Studies by Meroney (2009) employed experimental and computational approaches to study external pressure on buildings. Guha et al. (2009) studied characterization of flow through openings on the Texas Technology University Building. Computational simulation result obtained for internal pressure responses of the test shows that Helmholtz frequency matches the analytical solution.

III. Materials and Methods

1.0  Mathematical model

In recent times, there have been efforts in combining computer fluid dynamics and atmospheric model capabilities to monitor the effects on air flows of different terrain. This is what is required to simulate different flow patterns. Different research has been carried out on different buildings. Brown et. al. (2001) measured velocity distributions for two- and three-dimensional building array in wind tunnel. The purpose of the experiments was to provide high quality and spatial dense data which is used to evaluate the performance of computational fluid dynamic models. The atmospheric model was improved to simulate air flow around buildings under influence of mesoscale wind variations (Yamada,2003).

2.0  Methods of analysis

Various analytical and numerical approaches have been employed for resolving different types of partial differential equations subject to suitable boundary conditions including nonlinear equations. However, numerical methods have been preferred because of the difficulty and accuracy associated with analytical techniques.

2.0.1  Finite difference method

The finite difference method was exclusively used for many years to solve numerically, differential equations. However, when dealing with situations like flows at very high Renolds number, flows around arbitrarily shaped objects and strongly time dependent flows, there are short comings such as numerical instability, lack of accuracy and difficulties in properly treating boundary conditions for curved walls.

2.0.2  Finite element method

The Finite element method is the most interesting practical technique for finding approximate solutions to partial differential equations in engineering and science. FEM is used to solve a wide variety of problems and in many instances, the only viable method for obtaining solutions. While the FEM is built on a rich mathematical background, it is still one of the most practical numerical schemes yet devised for solving complex problems.

This method requires division of problem domain in many sub domains and each sub domain is called a finite element. Therefore, the problem domains consist of many finite element patches. One of the major advantages of finite element method is that a general-purpose computer program can be developed easily to analyses various kind of problems. Any shape can be handled with ease using this method. The procedures involve in FEM are as stated below:
- Discretization of the solution domain
- Selection of proper interpolation model
- Derivation of element stiffness matrices
- Assemblage of element equation to obtain overall equilibrium equation.
- Solution for the unknowns in the model equation.

A sequence of the approximate solutions is obtained as the element is reduced successively. If the underlisted conditions are satisfied, the sequence will converge to the exact solution.
- The field variable must be continuous.
- All uniform states of the field variable and its partial derivation in the functional must have representation in the interpolating polynomial.
- The field variable and its partial derivative must be continuous at the element boundary.

2.0.3  Finite volume method

The finite volume method is associated with special discretization of the domain and offers a more readily understood weighted residual for approximating the solution of the partial differential equation through local conservation. FVM is not associated with mathematical analysis as with FEM.

FVM is involved with mathematical analysis in relation to stability and convergence.

2.0.4  Computational fluid dynamics

One of the methods to estimate the wind loading acting on buildings is to measure the mean pressure and the root mean square pressure (pressure fluctuation) on the building envelope. The mean pressure and the root mean square pressure can be obtained by conducting wind tunnel test. However, it is costly to conduct a wind tunnel test. Computational fluid dynamic (CFD) is an alternative way to solve this problem. The mean pressure acting on building can be obtained using a turbulent model called k - ε model, but the estimation of pressure fluctuations needs to be relied on a model called Large Eddy Simulation (LES). Although the 3D LES can give a good analysis of the flow fields. To predict the wind-induced pressure fluctuations more efficiently, three main procedures are involved. Firstly, predict the mean flow quantities such as mean velocity flow field, turbulent kinetic energy (k) and turbulent energy dissipation (ε) using the modified k-ε model. A modified k-ε model is used to obtain a more accurate turbulent kinetic energy near the building. Secondly, generate a velocity fluctuation flow field that satisfies the mean turbulent quantities. Finally, solve the Poisson equation to predict the pressure fluctuation. The Poisson equation is derived from the incompressible momentum equations and continuity. This model is applicable for both 2D, and 3D simulation and it is believed that this model requires less computational effort comparing to the LES model.

2.0.5 Direct numerical simulation

Direct numerical simulation of the Navier stroke equation for a full range of turbulent motion for all scales (large down to dissipation) is the goal for numerical simulation of fluid flow. It is the most accurate way to model fluid flow numerically (Murakami, 1997). The only approximations made would be those necessary numerically and would be chosen to minimize discretization errors. When properly carried out, DNS results would be comparable in every way to quality experiment data (Ferziger, 1993). The main advantages are the clear definition of all conditions (initial, boundary and forcing) and the production of data for every single variable. However, from a practical viewpoint, only simple geometries and low Reynolds number will be modeled and while DNS is unsurpassed in its ability to predict turbulence, it is unlikely to become a practical engineering tool (Speziale, 1998).

However, basic computation using DNS provides very valuable information for verifying and revising turbulent models (Murakami, 1998).

2.0.6 Boundary conditions

Boundary conditions must be physically realistic. Hence, dependent on the geometry, materials, and the value of pertinent parameters. For this study, a flow over a low-rise building, the building has a solid boundary that is no slip boundary conditions.

The importance of the use of physical meaningful boundary conditions in numerical simulation cannot be over stressed, because if not properly defined can lead to error.

3.0 Governing equation

The governing equations for fluid flow are equations of rate and change in position to forces that cause(s) deformation. They define the property of fluid by the function of space and time (Fox and Mc Donald, 1994). It can be in the integral form (control volume) or differential form (point to point) forms. Differential forms will be employed in this research.

The governing equation for numerical simulation of wind pressure load on low rise building are the continuity equation which justified the assumption that air flow via wind effect could be treated as a continuous distribution of matter (Fox and Mc Donald, 1994). The momentum equation governs the rate of transfer and transport of air and the associated forces and the energy equation.

However, it is assumed that these equations are closed systems of nonlinear partial differential equations.

3.0.1 Mathematical formulation

Applying the fundamental laws of mechanics to a fluid gives the governing equations for a fluid. The conservation of mass equation is:

$$\frac{\partial \rho}{\partial t} + \nabla \cdot (\rho \vec{v}) = 0 \qquad (2)$$

and the conservation of momentum equation is:

$$\rho \frac{\partial \vec{v}}{\partial t} + \rho (\vec{v} \cdot \nabla) \vec{v} = -\nabla P + \rho \vec{g} + \nabla \cdot \tau_{ij} \qquad (3)$$

These equations along with the conservation of energy equation form a set of coupled, non-linear partial differential equations. It is not possible to solve these equations analytically for most engineering problems. However, it is possible to obtain approximate computer-based solutions to the governing equations for a variety of engineering problems. This is the subject matter of Computational Fluid Dynamics (CFD).

3.0.2 Prediction of mean flow quantities

In the current model, mean flow calculations are made using the standard K-ε model. The governing equations of the standard k- ε model are:

Continuity

$$\frac{\partial U_i}{\partial \chi_i} = 0 \qquad (4)$$

$$\frac{DU_i}{Dt} = -\frac{1}{\rho}\frac{\partial P}{\partial \chi_i} + \frac{\partial}{\partial X_j}\left[(v + v_t)(\frac{\partial U_j}{\partial x_i} + \frac{\partial U_j}{\partial x_i})\right] \qquad (5)$$

Turbulent kinetic energy

$$\frac{DK}{Dt} = \frac{1}{\rho}\frac{\partial}{\partial X_j}\left[\left(v + \frac{v_t}{\sigma_k}\right)\frac{\partial K}{\partial x_J}\right] + P_k - \varepsilon \qquad (6)$$

Energy dissipation

$$\frac{D\varepsilon}{Dt} = \frac{1}{\rho}\frac{\partial}{\partial X_j}\left[\left(v + \frac{v_t}{\sigma_k}\right)\frac{\partial \varepsilon}{\partial x_J}\right] + \frac{\varepsilon}{K}(C_1 P_k - C_2 \varepsilon) \qquad (7)$$

where the eddy viscosity $V_t$ is expressed as a function of the turbulent kinetic energy k, and the energy dissipation rate ε as

$$V_t = C_\mu \frac{K^2}{\varepsilon} \qquad (8)$$

In the above equations $P_k$ is given by

$$P_k = V_t S^2 \qquad (9)$$

$$v_t = C_\mu^* \frac{K^2}{\varepsilon}, \qquad C_\mu^* = C_\mu \frac{\Omega}{S}(\frac{\Omega}{S} < 1) \qquad (10)$$

$$v_t = C_\mu^* \frac{K^2}{\varepsilon}, \qquad C_\mu^* = C_\mu \frac{\Omega}{S}(\frac{\Omega}{S} \geqslant 1) \qquad (11)$$

Where;

$$S = \sqrt{\frac{1}{2}(\frac{\partial <u_i>}{\partial x_j} + \frac{\partial <x_j>}{\partial x_i})^2} \qquad (12)$$

$$\Omega = \sqrt{\frac{1}{2}(\frac{\partial <u_i>}{\partial x_j} - \frac{\partial <x_j>}{\partial x_i})^2} \qquad (13)$$

4.0 Applications of CFD

FLUENT, like other commercial CFD codes, offers a variety of boundary condition options such as velocity inlet, pressure inlet, pressure outlet, etc. It is very important to specify the proper boundary conditions to have a well-defined problem. The alternative to DNS found in most CFD packages (including FLUENT) is to solve the Reynolds Averaged Navier Stokes (RANS) equations. RANS equations govern the mean velocity and pressure because these quantities vary smoothly in space and time, they are much easier to solve.

In the determination of the profile, the software GAMBIT will be used to design the profile in which it will be meshed before it will be export to FLUENT.

IV. RESULT AND DISCUSSION

In the consideration of the numerical study of wind pressure load on low rise buildings under different terrain. Analysis was carried out using the computer software known as ANSYS Fluent 6.2 for the simulation and the governing equation is the Reynolds Average Navier Stoke (RANS) equation together with the k-e model.

The simulation was carried out for 3 different building roofs which are: the flat, gable and circular roof. Also, two different terrains were considered which are the flat and shallow escarpment.

The model of the building was developed in GAMBIT which is a modeling software that works with Fluent 6.2. A 2D model of three different roofs was used. They are flat roof model which is made up of front wall, rear wall and a roof; gable roof model which is made up of a front wall, rear wall, front roof and rear roof; and circular roof that consisted of a front wall, rear wall and roof.

The meshes were generated using quadrilateral cells of a known dimension. The inlet boundary specified a defined pressure outlet, and the model is specified as the walls. After the specification, it is then imported to the commercial software known as ANSYS Fluent for simulation. The simulation is pressure based in which the velocity formulation is absolute, and the viscous model is based on k-epsilon using standard wall function. The Fluent fluid material used is air (as wind) and the operating pressure is the atmospheric pressure.

The simulation was performed for the 3 different types of roofs at different velocities which are 12m/s, 15m/s and 20m/s in order to compare the effect of wind at various speeds. Also, this was analyzed for the flat and shallow escarpment terrain.

1.0 Flat Roof on a Shallow Escarpment

A flat roof is considered with a shallow escarpment for two different velocities of wind through the windward and leeward direction. This shows discrepancies in the velocity magnitude, total pressure, and the stream function.

Fig 9a & b shows the contours of total pressure around the roof section of the building. An area known as the stagnation zone is noticed on the top of the roof which shows that the static pressure drops to negative (shown by the blue color on the roof). It is evident that as the speed of wind increases, the stagnation zone also increases and can also be verified from Fig 10a & b which represent the contour at the leeward side of the building. The vortex formed at the leeward side of the building is proportional to the velocity of wind. The contour of stream function showing the path-lines, result in separation shear layer at the top of the roof as shown in Fig 11a & b. It is evident that increase in velocity of wind leads to development of vortices from the rolling up of shear layer. The contour for coefficient of pressure for a flat roof at a shallow escarpment is shown in Fig 12a & b.

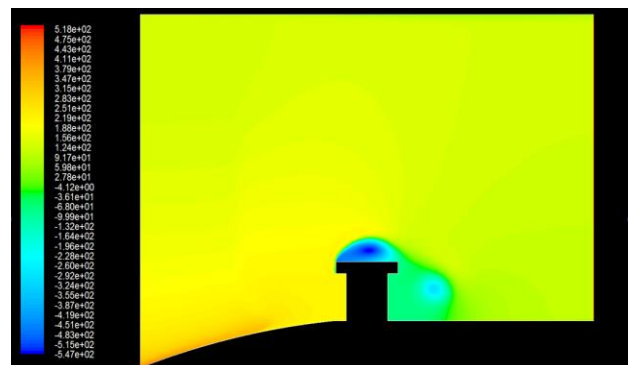

Figure 9a. The contour of total pressure around the roof section of flat roof model at wind velocity of 12m/s.

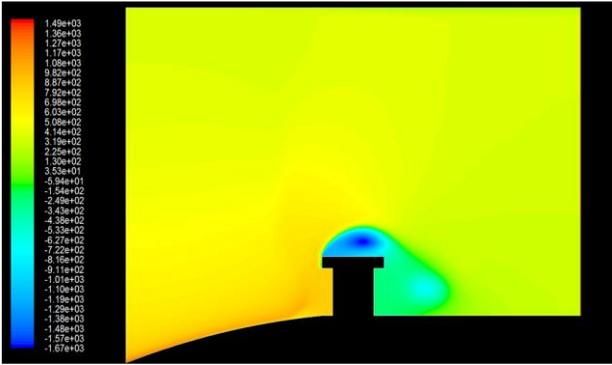

Figure 9b. The contour of total pressure around the roof section of flat roof model at wind velocity of 20m/s.

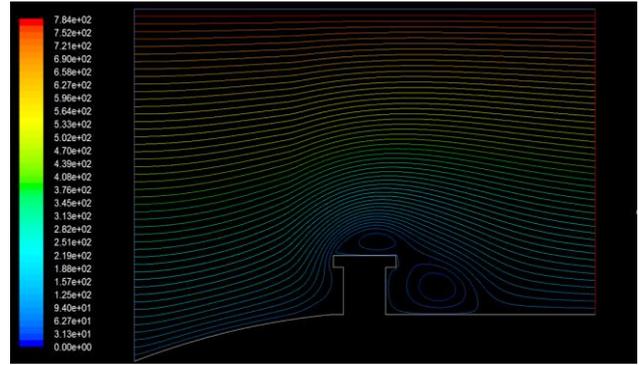

Figure 11b. The contours of the stream function around the building model at velocity of 20m/s.

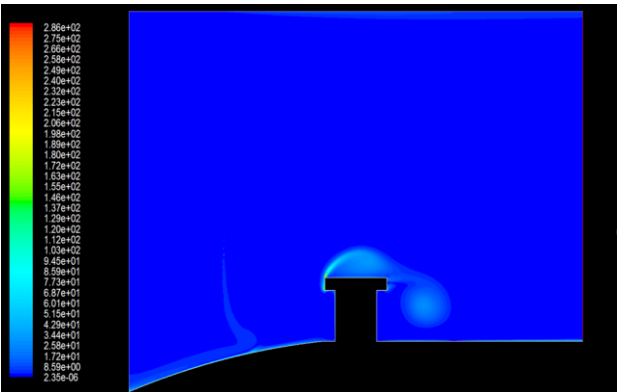

Figure 10a. The contours of vorticity magnitude at leeward side of the model at a wind speed of 12m/s.

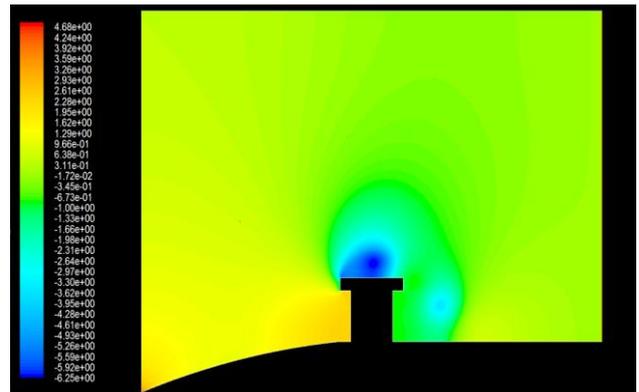

Figure 12a. The contours of coefficient of pressure for a flat roof model at a wind speed of 12m/s.

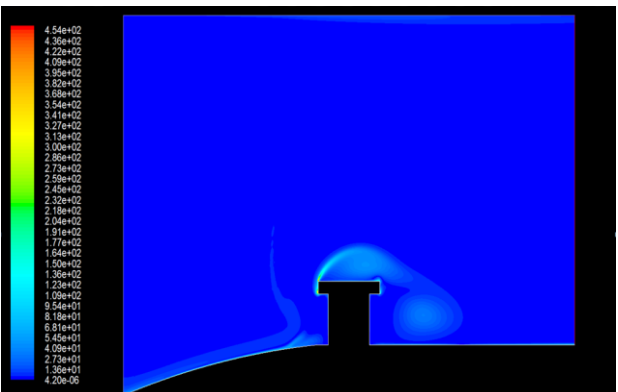

Figure 10b. The contours of vorticity magnitude at leeward side of the model at a wind speed of 20m/s.

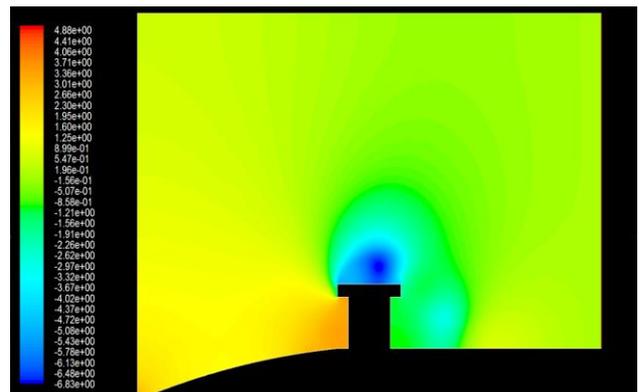

Figure 12b. The contours of coefficient of pressure for a flat roof model at a wind speed of 20m/s.

Fig 13a shows the coefficient of drag plotted against the flow time for a flat roof model under a shallow escarpment terrain. This shows that as the velocity increases, the flow time, and the coefficient of drag decreases.

Fig 13b shows the average weighted area of the model against the time step in which the velocity of wind increases as the average weighted area of the model increases at a constant time step.

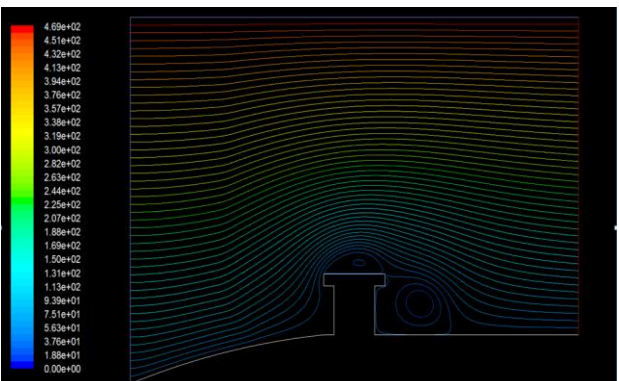

Figure 11a. The contours of the stream function around the building model at velocity of 12m/s.

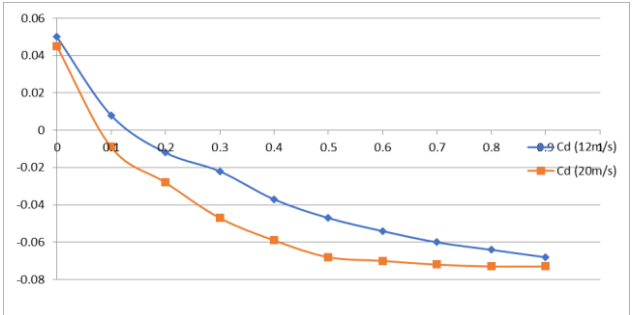

Figure 13a. The coefficient of drag against flow time for flat roof model at different velocity.

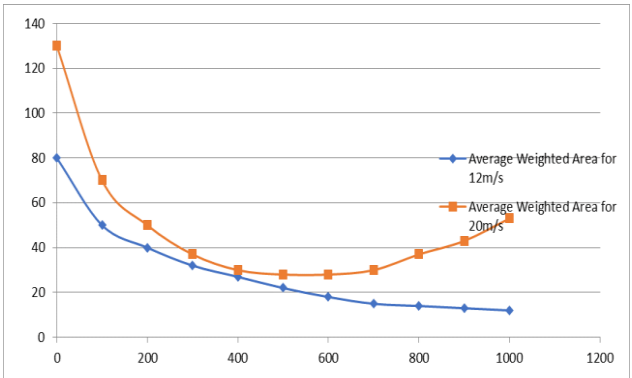

Figure 13b. The average weighted area against time step for flat roof model at different velocity.

2.0 Roof types on a Flat terrain

Fig 14 shows the coefficient of drag against flow time for the 3 different types of roofs at a constant velocity of 12m/s.

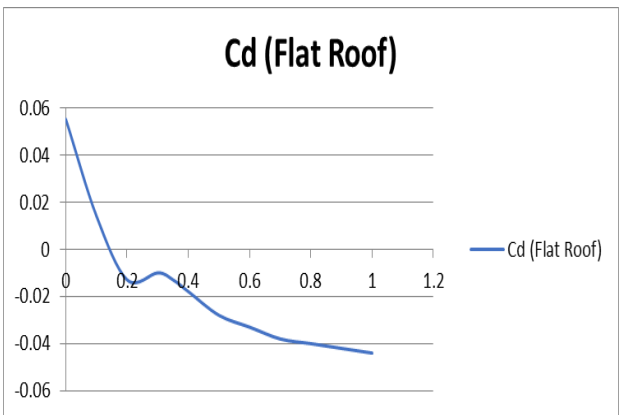

(a)

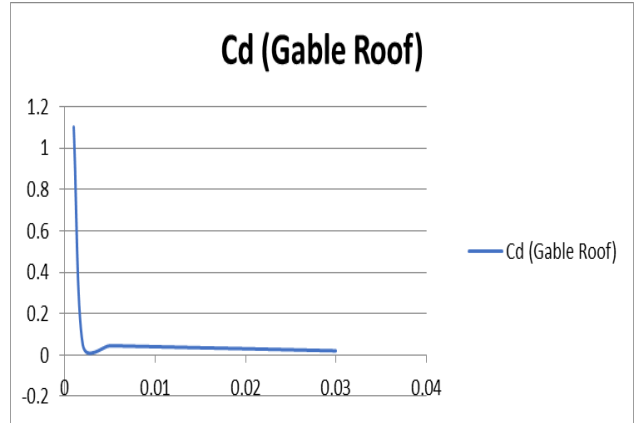

(b)

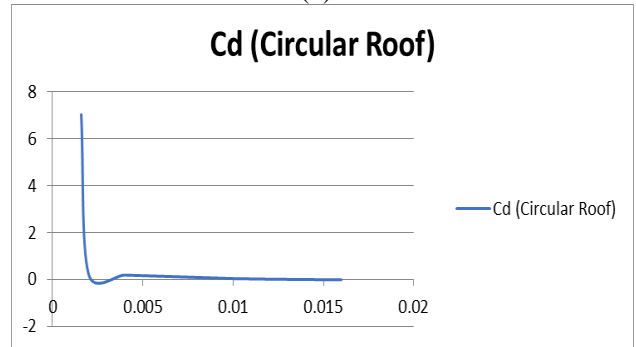

(c)

Figure 14. The coefficient of drag against flow time for the different roof model at velocity of 12m/s

Fig 15 shows the total pressure around the 3 different roofs of the model in which the stagnation zone is formed in the roof of the building which implies a negative pressure. It implies that the pressure decreases most for the flat roof, next to it is the gable roof followed by the circular roof.

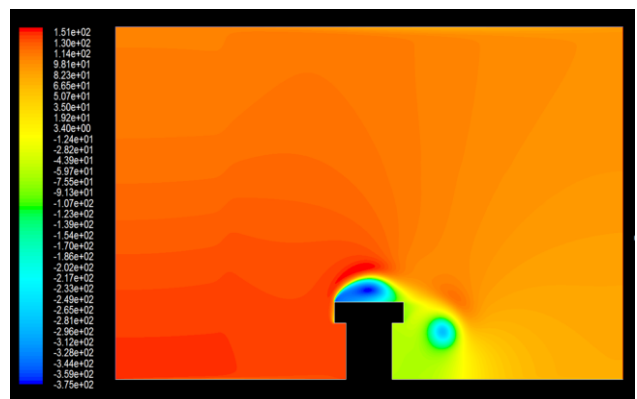

(a)

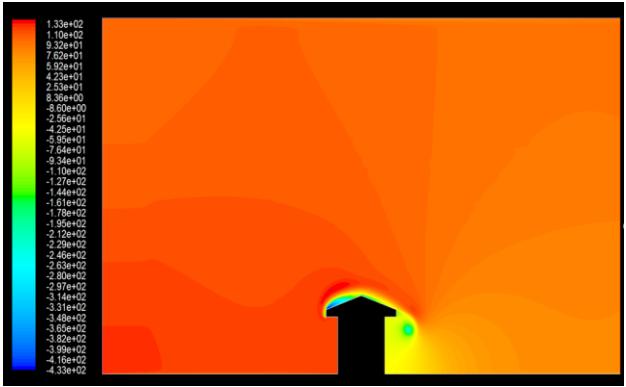

(b)

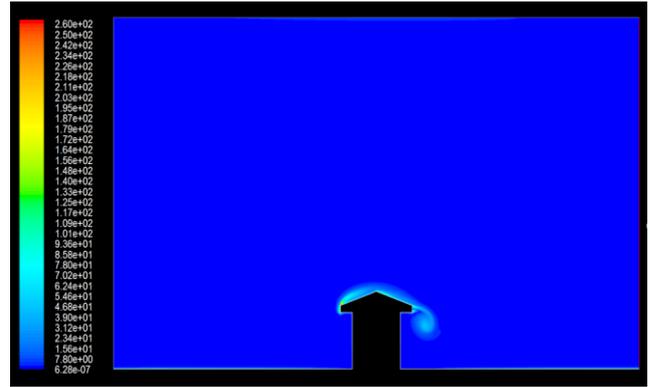

(b)

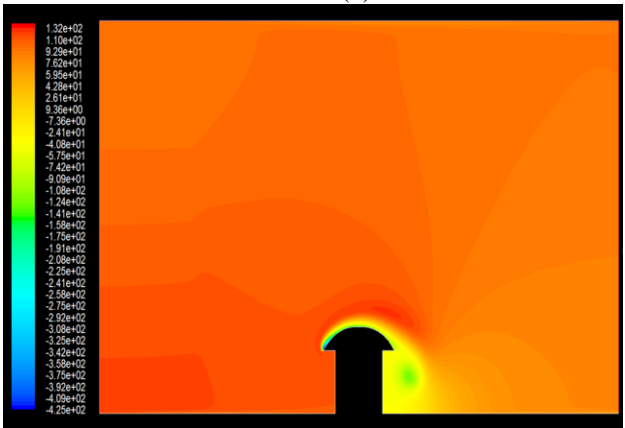

(c)

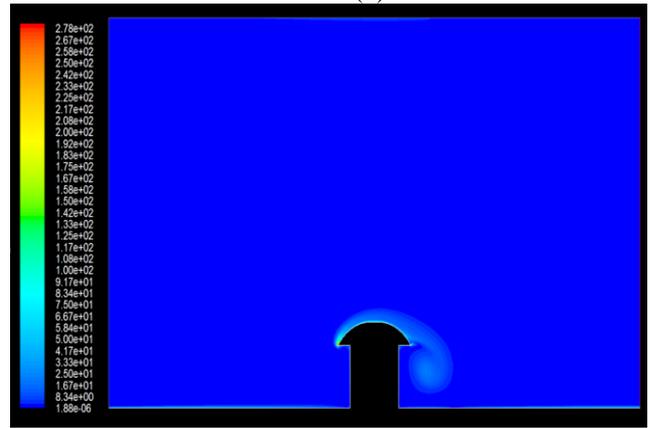

(c)

Figure 15. The contour of the total pressure for 3 different types of roofs at a velocity of 12m/s.

Figure 16. The contour of the vorticity magnitude for 3 different types of roofs at velocity of 12m/s.

Fig 16 shows the contour of vorticity magnitude around the 3 different roofs of the model in which the gable roof has the highest pressure, next to it is the circular and then the flat roof. The vortex formed is highest in the flat roof, then the circular and the gable.

Fig 17 shows the stream function for the 3 different roofs, this implies that the flat roof possesses the highest vortex formed then the circular and the gable roof. For the flat roof, the vortex increases over time which will lead to an obstruction of wind flow, which in turn causes damage to the roof.

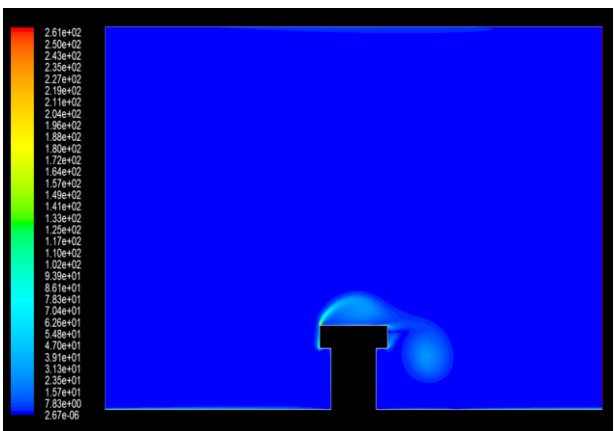

(a)

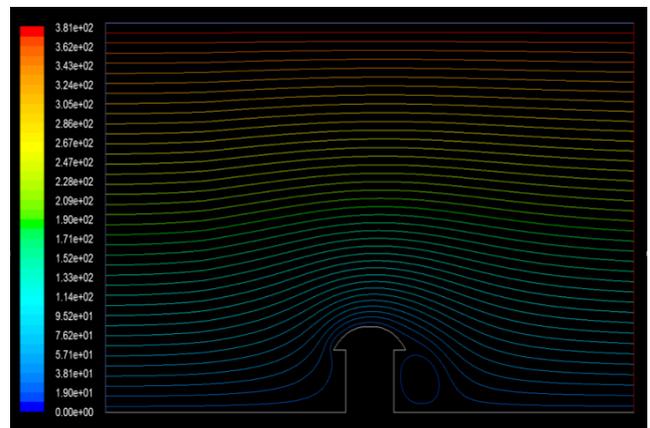

(a)

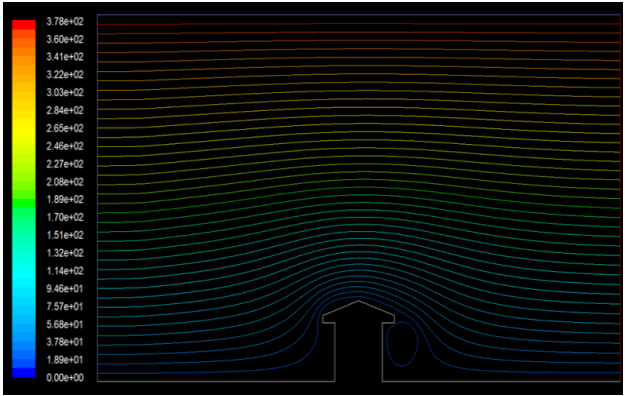
(b)

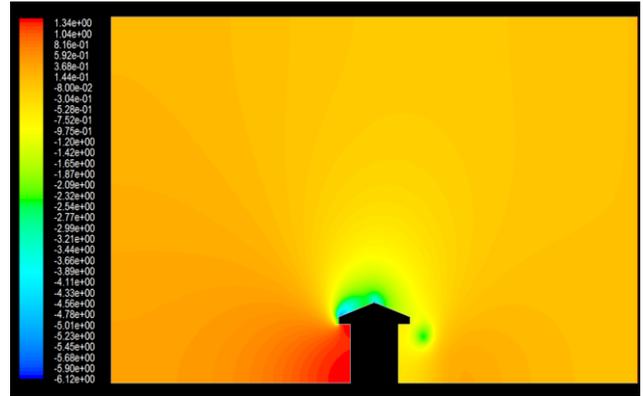
(b)

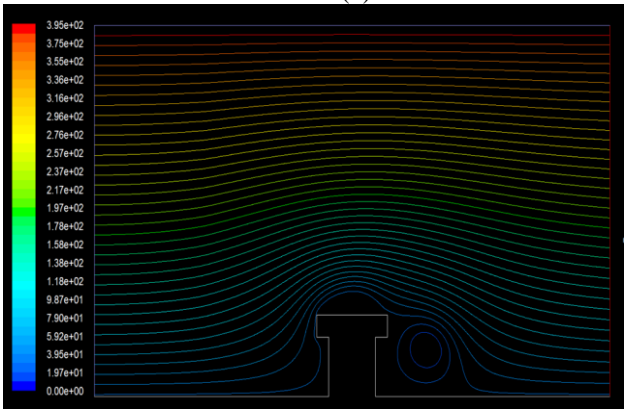
(c)

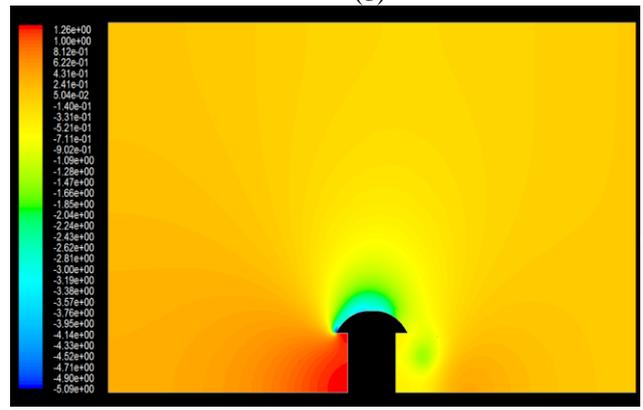
(c)

Figure 17. The stream functions for 3 different types of roofs at a velocity of 12m/s.

Figure 18. The contours of coefficient of pressure for the different roof model at a velocity of 12m/s

Fig 18 shows the coefficient of pressure for the 3 different types of roofs at a constant velocity of 12m/s.

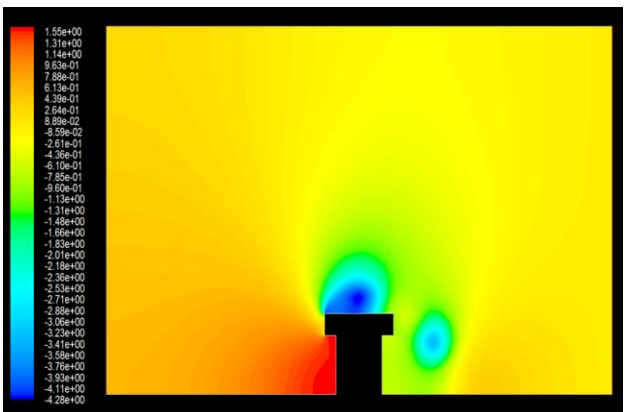
(a)

V. CONCLUSION

The numerical study of wind pressure load on low rise buildings under different terrain was carried out in which the coefficient of drag was obtained for the roof type considered at different velocities. The k-$\varepsilon$ model used which is pressure based is used to obtain the changes in the contours of the total pressure, vorticity magnitude and the stream function for different types of roofs at different terrain. This study has been able to identify the effect of high and low speed wind on a building and the best type of roof for construction. Also, the aerodynamic characteristics of wind around building walls and roofs.